\documentclass[preprint,prd,aps,showpacs,showkeys,nofootinbib]{revtex4}

\def\XXint#1#2#3{{\setbox0=\hbox{$#1{#2#3}{\int}$}
     \vcenter{\hbox{$#2#3$}}\kern-.5\wd0}}

\usepackage{graphicx}
\usepackage{bm}
\usepackage{amssymb}
\usepackage{amsmath}
\usepackage{euscript}

\begin{document}

\title{A quantum bound on the $1/f$ noise in semiconductors with a conical energy-momentum dispersion}

\author{Kirill~A.~Kazakov}

\affiliation{Department of Theoretical Physics,
Physics Faculty,\\
Moscow State University, $119991$, Moscow, Russian Federation}

\begin{abstract}
The quantum indeterminacy caused by non-commutativity of observables at different times sets a lower bound on the voltage noise power spectrum in any conducting material. This bound is calculated explicitly in the case of semiconductors with a conical energy-momentum dispersion of charge carriers. It possesses all characteristic properties of $1/f$ noise. Its momentum decomposition is found to be singular at zero particle momentum, a measurable consequence being a sharp peak in the noise magnitude at small charge carrier density. In application to monolayer graphene, this peak becomes M-shaped on account of a continuous transition from the electron to hole conductivity. A comparison with experimental data is made which demonstrates that the calculated power spectrum is close in magnitude and congruent to the observed.
\end{abstract}

\maketitle

\section{Introduction}\label{introduction}

It is well known that the spectra of voltage fluctuations in materials possessing free-like charge carriers are to a large degree universal at low frequencies: the power spectral density of fluctuations $S(f)\sim 1/f^{\gamma},$ where the frequency exponent $\gamma$ is around unity \cite{johnson, buckingham,bell1980,raychaudhuri}. This so-called $1/f$ noise has been detected down to frequencies as low as $f=10^{-6.3}$ Hz \cite{rollin1953,caloyannides}, with no sign of a low-frequency cutoff. There are many physical processes in conducting media which affect electrical conductivity at random and thereby produce voltage noise: charge carrier trapping, temperature fluctuations, generation-recombination of carriers in semiconductors, impurity migration, {\it etc.} The corresponding noise spectra are well understood, and their detection is a means to identify the underlying processes in specific materials. An eventual flattening at low frequencies and comparatively narrow frequency spans (three decades at most) are two most common features of these conventional spectra. As regards to $1/f$ noise, despite the lack of a conductance-fluctuation model that would explain altogether its amazing ubiquity, the absence of a low-frequency cutoff and the observed magnitudes of $1/f$ spectra, it is currently widely accepted that this noise, too, is generated by fluctuations in electrical conductance. On the other hand, it was shown recently that the quantum nature of charge carrier interaction with photons sets a lower bound on the spectral power of voltage fluctuations, and moreover, that this bound possesses all characteristic features of the observed $1/f$ noise \cite{kazakov1,kazakov2}. This bound was calculated explicitly in the case of charge carriers with a parabolic energy-momentum dispersion, and a comparison with $1/f$-noise in InGaAs quantum wells and high-temperature superconductors was made which led to the conclusion that the measured noise levels are only a few times as high as the bound established. The purpose of this Letter is to obtain a similar bound in another important case -- a conical energy-momentum dispersion. This case is of special interest in view of the fact that the bound found in Refs.~\cite{kazakov1,kazakov2} is inversely proportional to the charge carrier mass, whereas the conical dispersion implies that the particles are massless.

To recapitulate main points of the new approach, a lower bound on the power spectral density (power spectrum) of voltage fluctuations arises from the quantum indeterminacy caused by non-commutativity of observables at different times. This indeterminacy obstructs applicability of the Wiener-Khinchin theorem, which otherwise leads to a divergent total noise power whenever $S(f)\sim 1/f^{\gamma}$ with $\gamma \geqslant 1.$ It is by this reason that the quantum bound involves no low-frequency cutoff. In fact, it applies equally well at all frequencies of interest in $1/f$-noise studies.  Consider a (semi)conducting sample with a constant electric current through it, and let the voltage across the sample be measured by means of two voltage probes which may or may not coincide with the current leads. For simplicity, the probes are assumed pointlike, $\bm{x}_1,\bm{x}_2$ denoting their position. A voltage between them measured at time $t$ is the sum of a constant bias $U_0(\bm{x}_1,\bm{x}_2)$ and a fluctuation, or noise, $\Delta U(t,\bm{x}_1,\bm{x}_2).$ The results of Ref.~\cite{kazakov1} can be summarized as a recipe for calculating the lower bound, $S_F(f),$ of the voltage power spectrum, as follows. The voltage fluctuation $\Delta U(t,\bm{x}_1,\bm{x}_2)$ is an observable to which there corresponds a Hermitian (Heisenberg) operator $\widehat{\Delta U}(t)$ (for brevity, the arguments $\bm{x}_1,\bm{x}_2$ in $\widehat{\Delta U}$ are henceforth omitted). Define a correlation of the fluctuations, $S(t-t') = \left\langle\widehat{\Delta U}(t)\widehat{\Delta U}(t')\right\rangle,$ where angular brackets denote the quantum-mechanical average over a fixed state of the system ``conducting sample plus electromagnetic field.'' $S_F(f)$ is determined by the part of $S(t-t')$ which is antisymmetric under the interchange $t\leftrightarrow t'$ ($\omega = 2\pi f$):
\begin{eqnarray}\label{sigma}
S_F(f) = \left|\lim\limits_{t_m\to \infty}\left\{\int_{-t_m}^{t_m}{\rm d}\tau S(\tau)\sin(\omega\tau) - \frac{1}{t_m}\int_{-t_m}^{t_m}{\rm d}\tau |\tau|S(\tau) \sin(\omega\tau)\right\}\right|.
\end{eqnarray} As was demonstrated in Refs.~\cite{kazakov1,kazakov2}, the antisymmetric part of $S(t-t')$ admits a Fourier decomposition despite the presence of $1/f$-term in the power spectrum, and the limit under the modulus sign in Eq.~(\ref{sigma}) may exist even if $S(\tau)$ does not vanish for $\tau \to \infty,$ as is the case when $S(f)\sim 1/f$ at low frequencies.

The expectation value $\left\langle\widehat{\Delta U}(t)\widehat{\Delta U}(t')\right\rangle$ can be computed using the Schwinger-Keldysh technique \cite{schwinger,keldysh}, according to which it
is written as
\begin{eqnarray}\label{expectation2}
\left\langle\widehat{\Delta U}(t)\widehat{\Delta U}(t')\right\rangle = {\rm tr}\left(\hat{\rho}\,\EuScript{T}_C\widehat{\Delta u}^{(2)}(t)\widehat{\Delta u}^{(1)}(t')\exp\left\{-{\rm i}\int_{C}{\rm d}t\,\hat{w}(t)\right\}\right),
\end{eqnarray}
where $\widehat{\Delta u}(t) = \hat{a}_{0}(t,\bm{x}_1)-\hat{a}_{0}(t,\bm{x}_2) - U_0,$ with $\hat{a}_{0}$ the scalar electromagnetic potential, $\hat{\rho}$ is the system density matrix, $\hat{w}$ is the interaction Hamiltonian, and the lowercase letters denote operators in the interaction picture; the so-called Schwinger-Keldysh time contour $C$ runs from $t=-\infty$ to $t=+\infty$, and then back to $t=-\infty,$ with all time instants on the forward branch [designated with a superscript $(1)$] treated as being in the past with respect to any time instant on the backward branch [designated with a superscript $(2)$]; $\EuScript{T}_C$ denotes operator ordering along this contour.

Our goal is to determine how $S_F(f)$ is changed by switching from a parabolic to a conical energy-momentum dispersion of the charge carriers. Therefore, the right hand side of Eq.~(\ref{expectation2}) will be evaluated under conditions otherwise identical to those of Ref.~\cite{kazakov1}, namely, the charge carriers will be assumed non-relativistic unpolarized fermions, with all spin indices accordingly suppressed, and the finite-temperature effects due to photon heat bath neglected. Thus, the charge carrier energy $\varepsilon$ is related to its momentum $\bm{q}$ as
\begin{eqnarray}\label{dispersion}
\varepsilon(\bm{q}) = v_F |\bm{q}|,
\end{eqnarray} where $v_F$ is the Fermi velocity. The assumption that the charge carriers are non-relativistic means that $v_F\ll c,$ the speed of light in vacuum. The charge carrier Hamiltonian is taken in the simplest form obtained by replacing the ordinary derivatives with covariant ones. The interaction Hamiltonian thus reads \begin{eqnarray}\label{hamiltonian}
\hat{w}(t) = \int {\rm d}^3\bm{x}\hat{\phi}^{\dagger}(x)\left[\varepsilon\left(-{\rm i}\hbar\bm{\nabla} - \frac{e}{c}\bm{a}\right) - \varepsilon\left(-{\rm i}\hbar\bm{\nabla}\right) + e\hat{a}_0(x)\right]\hat{\phi}(x), \quad x = (ct,\bm{x}),
\end{eqnarray} where $\hat{\phi}(x)$ is the fermionic field, and
$${\bm a} = {\rm i}c\bm{E}\frac{{\rm e}^{{\rm i\lambda t}} - 1}{\lambda}\,, \quad \lambda \to 0,$$ is the vector potential of the electric field $\bm{E}$ established in the sample. The lowest-order contribution to the correlation function is quadratic with respect to $\bm{E}.$ Therefore, the relevant interaction vertices generated by $\hat{w}(t)$ include, in momentum space, a linear vertex
$$\hat{w}_1(t) = - \frac{v_F e}{c} \int\frac{{\rm d}^3 \bm{q}}{(2\pi)^3}(\bm{v},\bm{a})\hat{\phi}^{\dagger}(\bm{q})\hat{\phi}(\bm{q}),$$ where $\bm{v}=\partial\varepsilon(\bm{q})/\partial\bm{q}$ is the charge carrier velocity, and a vertex quadratic in $\bm{E},$ $$\hat{w}_2(t) = \frac{v_F e^2}{2c^2} \int\frac{{\rm d}^3 \bm{q}}{(2\pi)^3}\frac{[\bm{q},\bm{a}]^2}{|\bm{q}|^3}\hat{\phi}^{\dagger}(\bm{q})\hat{\phi}(\bm{q}),$$
$(\cdot,\cdot)$ and $[\cdot,\cdot]$ denoting the scalar and vector product, respectively.

\begin{figure}
\includegraphics[width=0.7\textwidth]{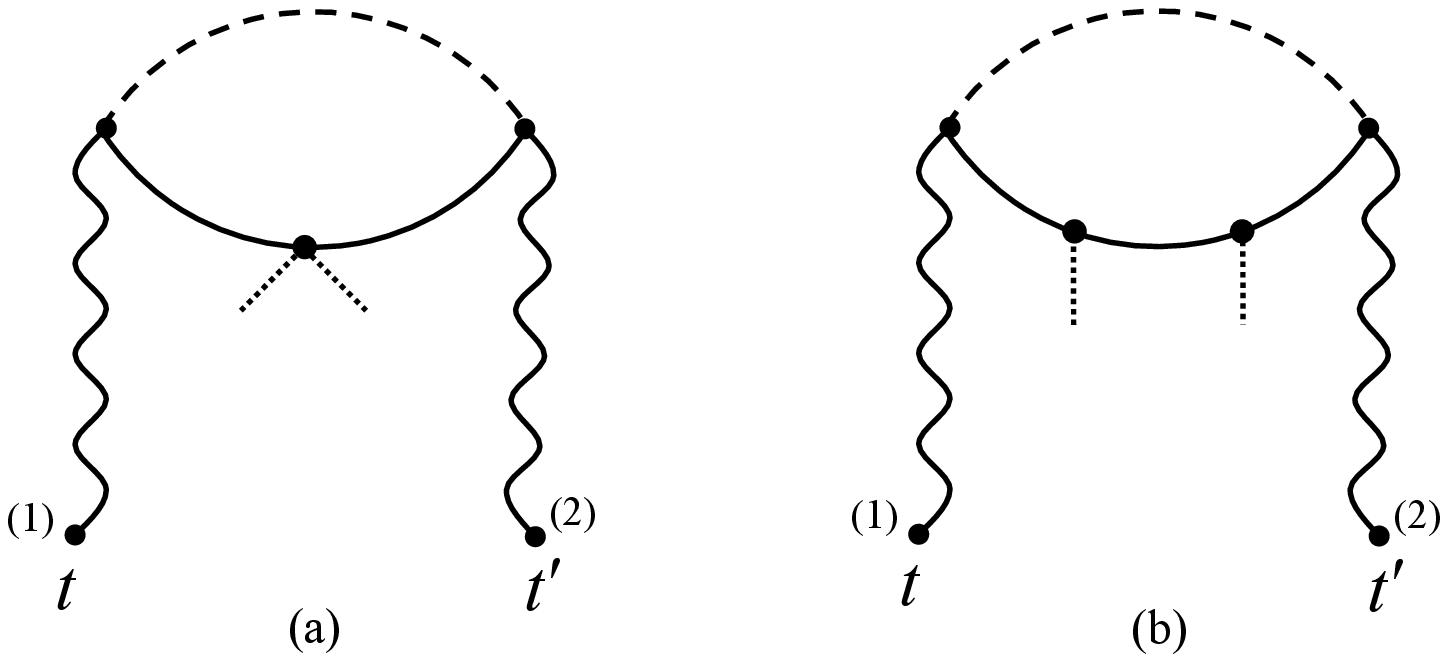}
\caption{Generic $O(\bm{E}^2)$-graphs representing the right hand side of Eq.~(\ref{expectation2}). Solid straight and wavy lines depict respectively the charge carrier and photon propagators, dashed line -- the charge carrier density matrix. Circles on the solid line with one or two dotted lines attached designate vertices $w_1$ and $w_2,$ respectively.}\label{fig1}
\end{figure}

\begin{figure}
\includegraphics[width=0.75\textwidth]{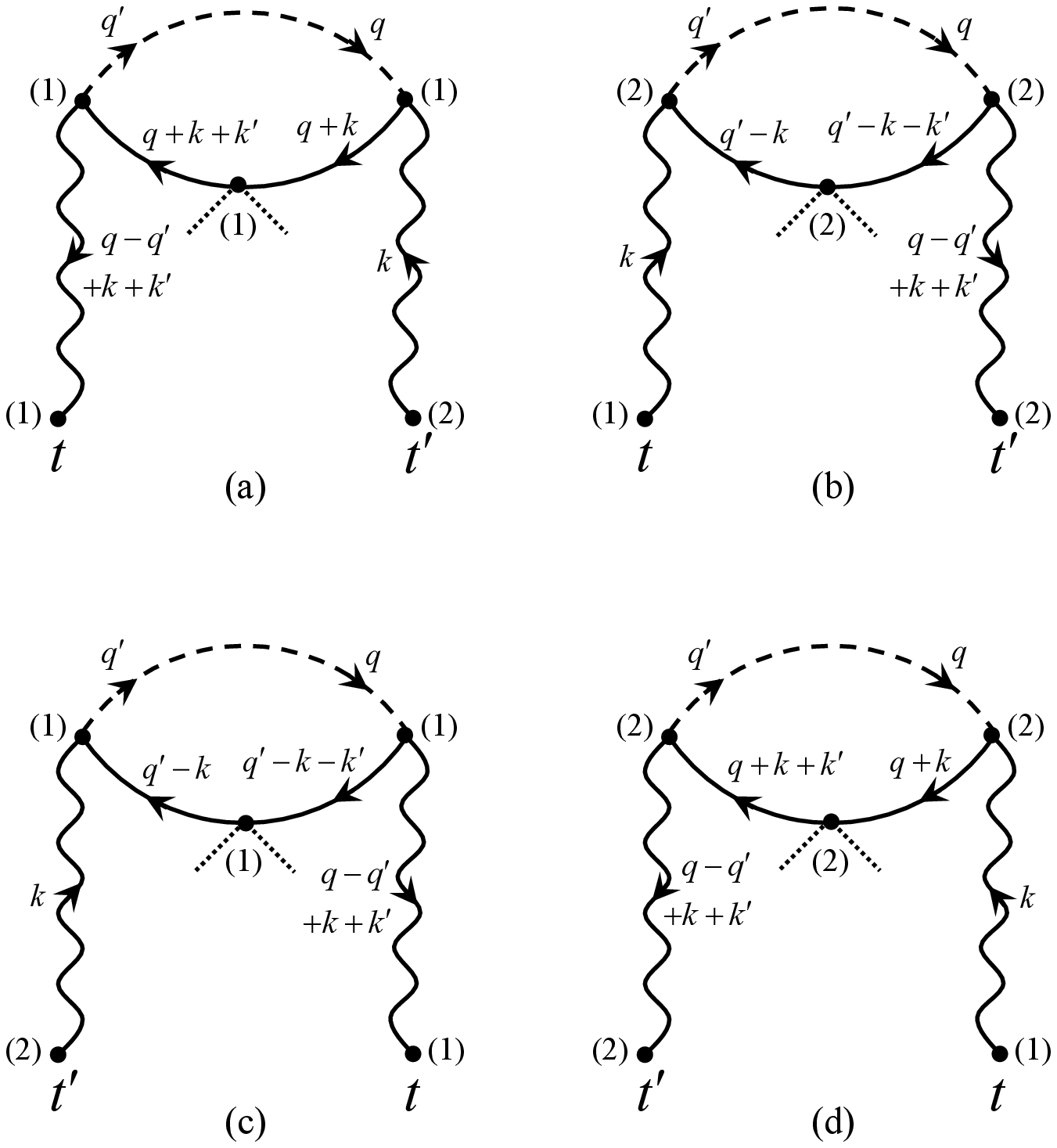}
\caption{Nonvanishing diagrams representing the graph Fig.~\ref{fig1}(b) in the Schwinger-Keldysh formalism. The arrows on the lines show the energy-momentum flow.}\label{fig2}
\end{figure}

The lowest-order contribution to the correlation function is represented by two generic graphs in Fig.~\ref{fig1}, while Fig.~\ref{fig2} is the set of nonvanishing Schwinger-Keldysh diagrams for the graph in Fig.~\ref{fig1}(a). The sum of these diagrams reads, in relativistic units $\hbar=c=1,$
 \begin{eqnarray}\label{longhand}
&&S(\tau) = \frac{{\rm i}v_F(4\pi e^2)^2}{2}\frac{\partial^2}{\partial\lambda\partial\lambda'}\int\frac{{\rm d}^4 k}{(2\pi)^4}\frac{{\rm d}^3 \bm{q}}{(2\pi)^3}\frac{{\rm d}^3 \bm{q}'}{(2\pi)^3}\varrho(\bm{q},\bm{q}')\left[{\rm e}^{{\rm i}\bm{k}\cdot(\bm{x}_1 - \bm{x}_2)} - 1\right] {\rm e}^{{\rm i}(\bm{q}-\bm{q}')\cdot\bm{x}_1 }\nonumber\\&&\times\left\{\left[{\rm e}^{-{\rm i}k^0\tau-{\rm i}k^{\prime0}t}G^{(11)}(q-q'+k+k')D^{(11)}(q+k+k')D^{(11)}(q+k)G^{(12)}(k)
\right.\right.\nonumber\\&& \left.\left. - {\rm e}^{{\rm i}k^0\tau-{\rm i}k^{\prime0}t'} G^{(22)}(q-q'+k+k')D^{(22)}(q+k+k')D^{(22)}(q+k)G^{(21)}(k)\right]
\frac{[\bm{q},\bm{E}]^2}{|\bm{q}|^3} \right.\nonumber\\&& \left. + \left[{\rm e}^{-{\rm i}k^0\tau-{\rm i}k^{\prime0}t}G^{(12)}(k)D^{(11)}(q'-k)D^{(11)}(q'-k-k')G^{(11)}(q-q'+k+k') \right.\right.\nonumber\\&& \left.\left. - {\rm e}^{{\rm i}k^0\tau-{\rm i}k^{\prime0}t'}G^{(21)}(k) D^{(22)}(q'-k)D^{(22)}(q'-k-k')G^{(22)}(q-q'+k+k')\right]
\left.\frac{[\bm{q}',\bm{E}]^2}{|\bm{q}'|^3}
\right\}\right|_{\lambda=\lambda'=0} \nonumber\\ &&+ (\bm{x}_1 \leftrightarrow\bm{x}_2),
\end{eqnarray} where $k' = (\lambda+\lambda',\bm{0}),$ $G$ and $D$ are the momentum-space Schwinger-Keldysh propagators of temporal photons and charge carriers, respectively,
\begin{eqnarray}\label{photonprop}
G^{(11)}(k) &=& \frac{{\rm i}}{k^2 + {\rm i}0} = [G^{(22)}(k)]^*, \quad G^{(12)}(k) = 2\pi\theta(-k^0)\delta(k^2), \quad G^{(21)}(k) = 2\pi \theta(k^0)\delta(k^2),\nonumber \\
D^{(11)}(q) &=& \frac{{\rm i}}{q^0 - \varepsilon(\bm{q}) + {\rm i}0} = [D^{(22)}(q)]^*, \quad k=(k^0,\bm{k}), \quad q=(\varepsilon(\bm{q}),\bm{q}),
\end{eqnarray} and the step function $\theta(k^0)=0$ for $k^0 \leqslant 0$, $\theta(k^0)=0$ for $k^0>0$; the photon momentum $\bm{k}$ is negligible compared to the charge carrier momenta $\bm{q},\bm{q}',$ and as such it is explicitly omitted in the vertex factors; since the photon heat bath effects are neglected, $G$'s are purely vacuum; accordingly, the system density matrix $\hat{\rho}$ is reduced to that of the charge carriers, $\varrho(\bm{q},\bm{q}')$ denoting its momentum-space representation; the latter is normalized by
$$\int\frac{{\rm d}^3 \bm{q}}{(2\pi)^3}\varrho(\bm{q},\bm{q}) = 1.$$
At last, ``$+ (\bm{x}_1 \leftrightarrow\bm{x}_2)$'' means that the preceding expression is to be added with $\bm{x}_1$ and $\bm{x}_2$ interchanged.

At low photon frequencies $k^0,$ the charge carrier propagators are near their poles, and therefore, the leading low-frequency term results from $\lambda$-differentiations of the factors $D^{(11)}(q+k+k'),$ $D^{(11)}(q'-k-k'),$ {\it etc.} To extract this term, we expand $\varepsilon(\bm{q+k}) \approx \varepsilon(\bm{q}) + (\bm{v},\bm{k})$ in the denominators and recall that $|\bm{v}| \ll 1,$ $|\bm{k}| = |k^0|,$ hence $|(\bm{v},\bm{k})|\ll |k^0|.$ The propagators are thus simplified as
\begin{eqnarray}\label{simplified}
D^{(11)}(q+k+k') &=& \frac{{\rm i}}{q^0 + k^0 + \lambda+\lambda' - \varepsilon(\bm{q+k})}\approx \frac{{\rm i}}{k^0+\lambda+\lambda'}, \nonumber\\ D^{(11)}(q'-k-k') &\approx& - \frac{{\rm i}}{k^0+\lambda+\lambda'},
\end{eqnarray}
{\it etc}. Substituting these expressions into Eq.~(\ref{longhand}) readily shows that the four diagrams give rise to pairwise equal contributions to the leading term (the pairs are enclosed in square brackets). Expanding also ${\rm e}^{{\rm i}\bm{k}\cdot(\bm{x}_1 - \bm{x}_2)}$ to the second order, and performing integration over $\bm{k}$ yields
\begin{eqnarray}\label{lowest}
S(\tau) =&& \frac{8v_Fe^4(\bm{x}_1 - \bm{x}_2)^2}{3}\int_{0}^{\infty}{\rm d}k^0\frac{{\rm e}^{{\rm i}k^0\tau}}{k^0}\int\frac{{\rm d}^3 \bm{q}}{(2\pi)^3}\frac{{\rm d}^3 \bm{q}'}{(2\pi)^3}\nonumber\\&&\times\varrho(\bm{q},\bm{q}')
\left(\frac{[\bm{q},\bm{E}]^2}{|\bm{q}|^3} + \frac{[\bm{q}',\bm{E}]^2}{|\bm{q}'|^3}\right)\frac{{\rm e}^{{\rm i}(\bm{q}-\bm{q}')\cdot\bm{x}_1 } + {\rm e}^{{\rm i}(\bm{q}-\bm{q}')\cdot\bm{x}_2 }}{(\bm{q} - \bm{q}')^2}\,.
\end{eqnarray} It is now easy to see that similar transformations of the Schwinger-Keldysh diagrams representing the graph in Fig.~\ref{fig1}(b) yield expressions which pairwise cancel each other (this is because these diagrams involve an odd number of the charge carriers propagators, which makes the sign of their contributions alternate). The leading low-frequency term of the spectral density of $S(\tau)$ is thus linear with respect to the Fermi velocity, but its dependence on the charge carrier momentum turns out to be more complicated than in the case of massive particles. To evaluate this dependence, it is convenient to express the charge carrier density matrix via the mixed position-momentum distribution function $R(\bm{r},\bm{Q}),$
\begin{eqnarray}\label{mixed}
\varrho\left(\bm{Q}-\frac{\bm{p}}{2},\bm{Q}+\frac{\bm{p}}{2}\right) = \frac{1}{\Omega}\int_{\Omega} {\rm d}^3\bm{r}{\rm e}^{{\rm i}\bm{p}\cdot\bm{r}}R(\bm{r},\bm{Q}),
\end{eqnarray} where $\Omega$ is the sample volume. $R(\bm{r},\bm{Q})$ vanishes for $\bm{r}$ outside of the sample. It follows that $(\bm{q}' - \bm{q}) = \bm{p}$ is of the order of the inverse linear sample size which in practice is much larger than the lattice constant. Therefore,  $|\bm{p}| \ll |\bm{q}|$ for all relevant charge carrier momenta, and so one can set $\bm{q}' \approx \bm{q} \approx \bm{Q}$ in the vertex factors. The integral over $\bm{p}$ in Eq.~(\ref{lowest}) is then just a Fourier decomposition of the Coulomb potential. Finally, the power spectrum of voltage fluctuations is found by substituting an odd in $\tau$ part of $S(\tau)$ into Eq.~(\ref{sigma}). Restoring ordinary units, we obtain
\begin{eqnarray}\label{sf1}
S_F(f) =&& \frac{2v_Fe^4\Delta\bm{x}^2}{3\pi \hbar c^3 |f|}\frac{1}{\Omega}\int_{\Omega} {\rm d}^3\bm{r}\left(\frac{1}{|\bm{r}-\bm{x}_1|} + \frac{1}{|\bm{r}-\bm{x}_2|}\right)\int\frac{{\rm d}^3 \bm{Q}}{(2\pi)^3}R(\bm{r},\bm{Q})\frac{[\bm{Q},\bm{E}]^2}{|\bm{Q}|^3}\,,
\end{eqnarray} where $\Delta\bm{x}\equiv\bm{x}_1 - \bm{x}_2.$ In the practically important case of macroscopically homogeneous sample, $R(\bm{r},\bm{Q}) = R(\bm{Q}),$ this expression further factorizes:
\begin{eqnarray}\label{sf2}
S_F(f) = \frac{2v_Fe^4\Delta\bm{x}^2}{\pi\hbar c^3 |f|}g\int\frac{{\rm d}^3 \bm{Q}}{(2\pi)^3} R(\bm{Q})\frac{[\bm{Q},\bm{E}]^2}{|\bm{Q}|^3}\,,
\end{eqnarray} where $$g = \frac{1}{3\Omega}\int_{\Omega} {\rm d}^3\bm{r}\left(\frac{1}{|\bm{r}-\bm{x}_1|} + \frac{1}{|\bm{r}-\bm{x}_2|}\right)$$ is a purely geometric factor (it is defined here as in the massive case).

It is seen that the power spectrum essentially depends on the momentum distribution of massless charge carriers, in contrast to the case of parabolic energy-momentum dispersion where it is independent of $R(\bm{Q}).$ Most notably, the integrand in Eq.~(\ref{sf2}) turns out to be singular for $\bm{Q}\to 0$: the factor $[\bm{Q},\bm{E}]^2/|\bm{Q}|^3 = O(1/|\bm{Q}|).$ Possible observational consequences of this singularity are naturally expected to be especially pronounced in low-dimensional systems and/or at small charge carrier densities, $n,$ wherein the characteristic charge carrier momentum is also small. This suggests consideration of $S_F$ as a function of $n.$

Let the sample be a thin semiconducting layer deposited on a dielectric substrate that allows controlling the charge carrier density in the sample. The sample thickness is supposed to be so small as to admit a two-dimensional treatment of the charge carrier motion. Assuming that the distribution of charge carrier momentum is isotropic in the sample plane, and that the voltage probes are aligned parallel to the electric field (as is usually the case in $1/f$ experiments), so that $\bm{E}^2\Delta\bm{x}^2 = U^2_0,$ Eq.~(\ref{sf2}) takes the form
\begin{eqnarray}\label{sf3}
S_F(f) = \frac{\varkappa U^2_0}{|f|}\,, \quad \varkappa =  \frac{v_Fe^4 g}{\pi\hbar c^3}\int\frac{{\rm d}^2 \bm{Q}}{(2\pi\hbar)^2} \frac{R(\bm{Q})}{|\bm{Q}|}\,.
\end{eqnarray} As defined by Eq.~(\ref{mixed}), $R(\bm{Q})$ is normalized in the two-dimensional case according to
\begin{eqnarray}\label{normalization}
\int\frac{{\rm d}^2 \bm{Q}}{(2\pi \hbar)^2}R(\bm{Q}) = 1.
\end{eqnarray} Therefore, on neglecting the effect of particle interactions on $R(\bm{Q})$, it can be written as $R(\bm{Q})=2N(\varepsilon(\bm{Q}))/n,$ where $N(\varepsilon)=[\exp(\varepsilon-\mu)/T+1]^{-1}$ is the Fermi--Dirac distribution, and $n$ has the meaning of surface density of charge carriers (the factor of two counts the electron spin states). Substitution of this relation in Eqs.~(\ref{sf3}), (\ref{normalization}) yields the reduced power spectrum $\varkappa$ and $n$ as functions of the chemical potential $\mu$:
\begin{eqnarray}\label{sf4}
\varkappa = \frac{2v_Fe^4 g}{\pi\hbar c^3n}\int\frac{{\rm d}^2 \bm{Q}}{(2\pi\hbar)^2} \frac{N(\varepsilon(\bm{Q}))}{|\bm{Q}|}\,, \quad n = 2\int\frac{{\rm d}^2 \bm{Q}}{(2\pi \hbar)^2}N(\varepsilon(\bm{Q})).
\end{eqnarray}

\begin{figure}
\includegraphics[width=0.55\textwidth]{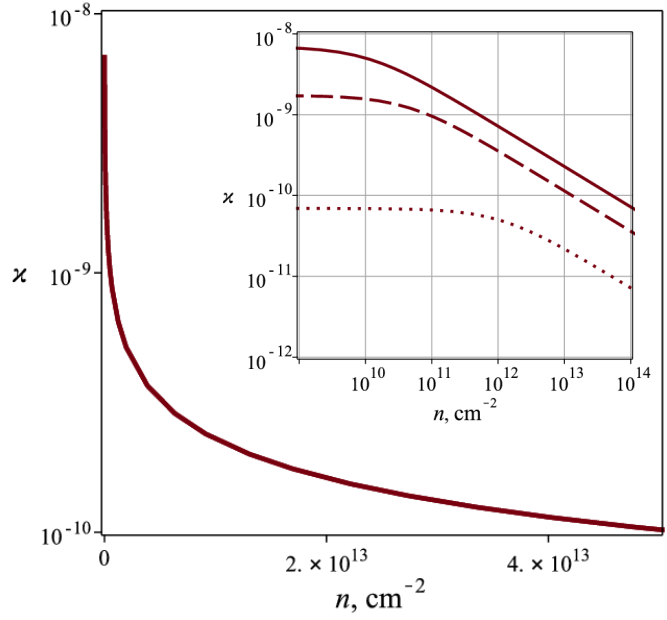}
\caption{Reduced power spectrum ($\varkappa$) versus charge carrier density ($n$) as given by Eq.~(\ref{sf4}) in a square sample with $g \approx 1.6\times 10^4\,$cm$^{-1}.$ $T=100\,$K, $v_F = 10^8\,$cm/s. Inset: log-log plot of the same function for a fixed $T=100\,$K and various $v_F = 10^7\,$cm/s (dotted line), $5\times 10^7\,$cm/s (dashed) and $10^8\,$cm/s (solid).}\label{fig3}
\end{figure}

Figure~\ref{fig3} is a parametric plot of the function $\varkappa(n),$  $\mu$ being the parameter, in the case of a square $1\times 1\,$$\mu$m$^2$-sized sample, with the voltage probes placed in the middle of its opposite sides. We see that the singularity at $\bm{Q}=0$ in the momentum decomposition of $\varkappa$ shows itself as a sharp increase of the noise magnitude at small charge carrier density. As the inset demonstrates, $\varkappa$ tends to a finite limit as $n\to 0,$ which is readily found to be
\begin{eqnarray}\label{maximum}
\varkappa|_{n=0} = \frac{v^2_Fe^4 g}{\pi\hbar c^3T}\,.
\end{eqnarray}
An asymptotic of $\varkappa(n)$ in the opposite case of large $n$ is also easy to determine analytically. For $n$ so large that $\mu\gg T,$ one has $$n \simeq \frac{\mu^2}{2\pi\hbar^2}\,, \quad \varkappa \simeq \frac{v_Fe^4g\mu}{\pi^2\hbar^3 c^3n}\,,$$ and therefore,
$$\varkappa \simeq\frac{v_Fe^4g}{\pi^{3/2}\hbar^2 c^3}\sqrt{\frac{2}{n}}\,.$$ This asymptotic is thus independent of temperature, and in fact, of the particular structure of   $N(\varepsilon)$ at $\varepsilon\sim \mu.$

In essence, the considered physical conditions are realized in the field-effect transistors based on single-layer graphene. The sample in this case is a planar monolayer of carbon atoms arranged in a hexagonal lattice \cite{geim2009}. The energy-momentum dispersion is mainly determined by the electron hopping between the two triangular sublattices, and is conical near the so-called Dirac points which are two corners of the graphene Brillouin zone \cite{wallace1947}. Near these points, the energy spectrum is symmetric with respect to zero energy, $\varepsilon(\bm{q})=\pm v_F|\bm{q}|,$ Dirac point being at $\bm{q}=0,$ though the symmetry is broken away the point by the next-to-nearest electron hopping. Thus, the charge carriers are two copies of electrons ($\varepsilon>0$) and holes ($\varepsilon<0$), all characterized by the same value of $v_F\approx 10^8\,$cm/s. The approximations made above imply disregard of the effects related to the electron spin (except for the number of states counting) and to the pseudo-spin (except for the energy-momentum dispersion). In view of this, Eq.~(\ref{sf4}) is to be considered only as an order-of-magnitude estimate of the quantum bound on $1/f$-noise in graphene. On the other hand, there is another, more general source of inaccuracy that makes exactness of calculations hard to improve. The point is that there is a natural interplay between the power spectrum magnitude and the value of the frequency exponent, $\gamma.$ As $\gamma$ deviates from unity, $\varkappa$ in Eq.~(\ref{sf4}) becomes dimensional, $\varkappa \to \varkappa (f_*)^{\delta},$ where $\delta \equiv \gamma - 1$ and $f_*$ is a frequency parameter which is on the order of the inverse characteristic time of the physical process responsible for the deviation. One such process is the charge carrier--phonon interaction, wherein $f_*$ is the ratio of the sound speed to the lattice constant \cite{kazakov2}. This parameter is a large number when measured in hertz, $f_* \approx 10^{13},$ so that even a comparatively small deviation of $\gamma$ can noticeably change the magnitude of $S(f).$ The available experimental data suggest that $\gamma$ is close to unity in graphene, but it is difficult to both calculate and measure accurately. The measurement error depends largely on the frequency span, and for $2-2.5$ decades commonly found in the $1/f$-studies in graphene, it is not less than $0.05.$ This error in $\gamma$ brings in a factor of $10^{13 \cdot 0.05}\approx 5$ in $\varkappa.$

\begin{figure}
\includegraphics[width=0.47\textwidth]{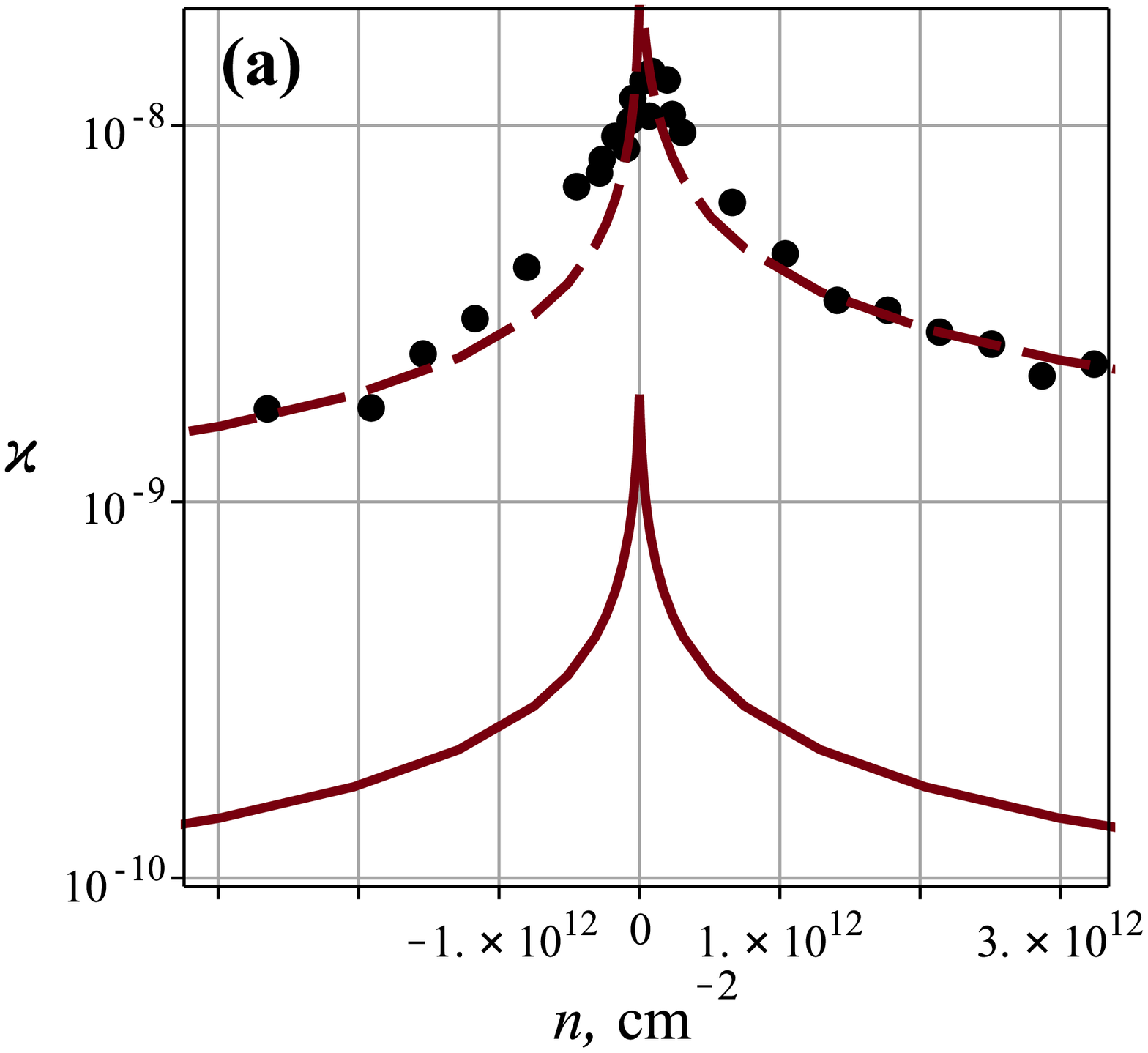}
\includegraphics[width=0.49\textwidth]{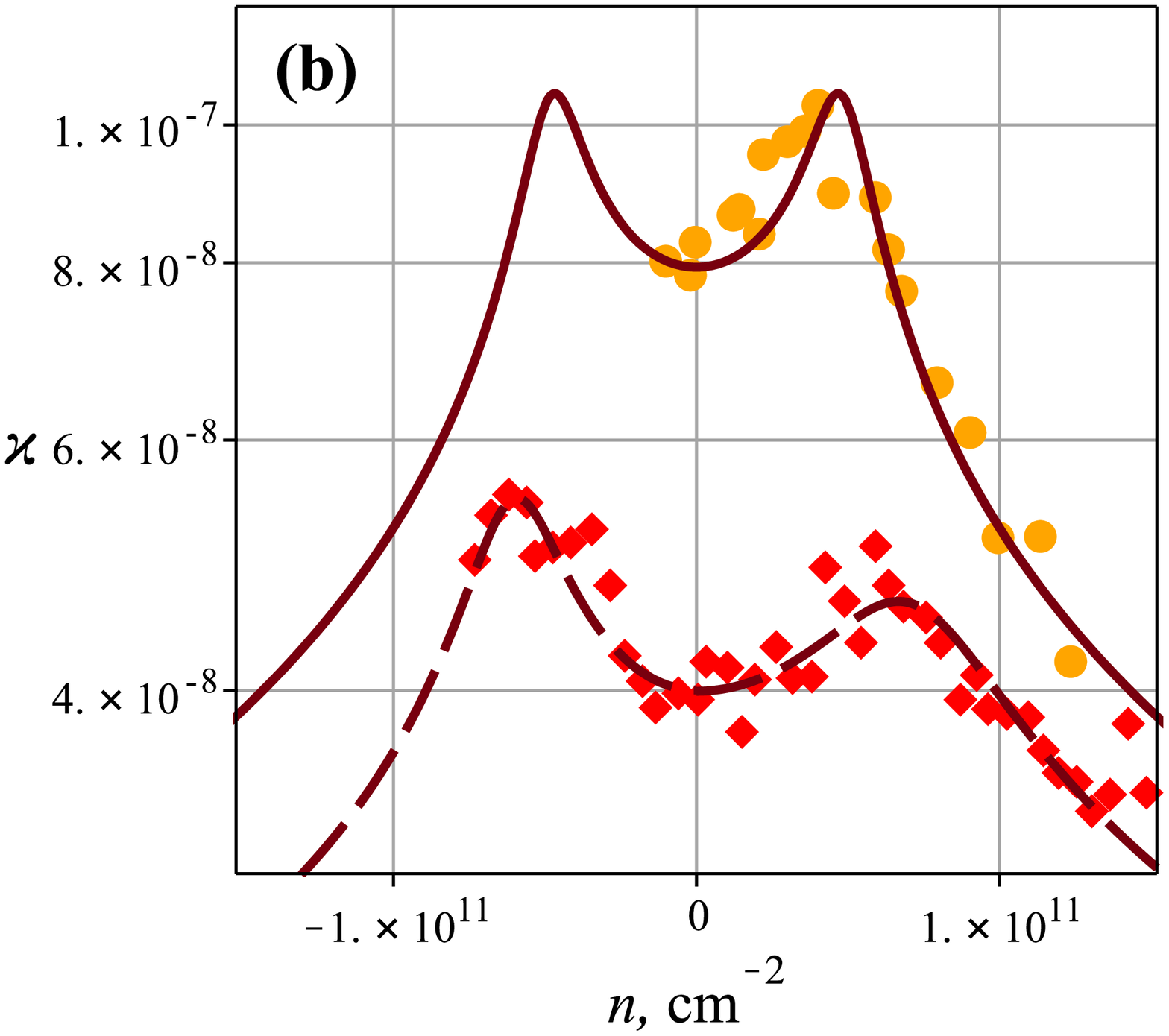}
\caption{(a) Reduced power spectrum versus charge carrier density as measured in Ref.~\cite{pal2011} in a field-effect device SLG1 (marks), and calculated according to Eq.~(\ref{sf4}) (solid line). Dashed line is a vertical shift of the solid. (b) Same in a suspended graphene flake at $T=300\,$K (solid line, circles), and at $T=200\,$K (dashed line, diamonds). $n>0$ ($n<0$) is the electron (hole) doping.}\label{fig4}
\end{figure}

Still, this ambiguity in the magnitude does not hinder comparison with the experimental data of the {\it trends} predicted by Eq.~(\ref{sf4}), in particular, the dependence of power spectrum on the charge carrier density. To this end, we use the results of Ref.~\cite{pal2011} where this dependence was measured in various graphene devices including graphene monolayer flakes deposited on SiO$_2$ substrates. Figure \ref{fig4}(a) compares the reduced power spectrum measured in a sample of size $3.5\times 4.5\,$$\mu$m$^2$ (marks, as read off from figure 1(d) of Ref.~\cite{pal2011}) with the calculated according to Eq.~(\ref{sf4}) wherein $n$ is doubled to account for the two Dirac points (solid line). The positive and negative $n$s correspond, respectively, to the electron and hole doping. Dashed curve is obtained by a vertical shift of the solid curve to match the experimental data. The right (left) wing of the curve is shifted by multiplying the theoretical $\varkappa$ for $\gamma=1$ with a factor of $15$ ($10$). Should this factor be a result of deviation of the frequency exponent from unity, the two numbers would correspond to $\delta = 0.09$ and $\delta=0.08$ (unfortunately, the comparison cannot be pursued in this direction, as Ref.~\cite{pal2011} does not specify values of $\gamma$).

Strictly speaking, the peak value (\ref{maximum}) is practically unattainable because of fluctuations in $n$ caused by disorder and charged impurities in the substrate. The corresponding characteristic density fluctuation in graphene, $\delta n \sim 10^{11}\,$cm$^{-2}$. But since $\varkappa$ slowly varies for $n\lesssim 10^{11}$ (Cf. Fig.~\ref{fig3}), $\varkappa|_{n=0}$ may well serve as an estimate of the $\varkappa$ maximum. There is also another important scale of electron density, $n_{\rm i},$ below which it becomes inhomogeneous because of occurrence of hole ``puddles'' within the sample, which grow and coalesce with decreasing electron density until they fill up the whole sample at $n=-n_{\rm i}$ \cite{hwang2007}. $n \in (-n_{\rm i},n_{\rm i})$ is thus a region of mixed electron-hole conductivity, the average electron (hole) density vanishing at its left (right) end. Under such circumstances, $R(\bm{r},\bm{Q})$ in Eq.~(\ref{sf1}) does depend on $\bm{r}$ within the sample, but if the puddles are distributed approximately uniformly throughout, Eq.~(\ref{sf4}) is still valid. Then, to take into account the vanishing of average electron/hole density at $n=-/+n_{\rm i},$ the right (left) wing of the curve $\varkappa(n)$ is to be shifted to the left (to the right) by the amount of $n_{\rm i};$ density fluctuations are allowed for by smearing the result on the scale $\delta n.$ This produces a curve with an M-shaped peak as drawn by a solid line in Fig.~\ref{fig4}(b), and this structure should be possible to experimentally resolve in samples with not-to-large $\delta n.$ On the other hand, M-shaped $\varkappa$s in graphene have been reported by a number of experimental studies \cite{lin2008,pal2009,heller2010,zhang2011,pal2011,kaverzin2012}; in particular, the authors of Ref.~\cite{pal2011} observed it in a device having the graphene flake partially suspended. This suspension reduces the disorder induced by the substrate, damping thereby fluctuations in the charge carrier density. The calculated dependence $\varkappa(n)$ is compared  with the experimental result \cite{pal2011} in Fig.~\ref{fig4}(b). The theoretical curves are drawn using the value $n_{\rm i}=5\times 10^{10}\,$cm$^{-2}$ as estimated in Ref.~\cite{pal2011}, and $\delta n$ on the order of $10^{11}\,$cm$^{-2}$ chosen so as to obtain best fits to the experimental data (the latter is read off from figure 2(c3) of Ref.~\cite{pal2011} where the sets of marks for different temperatures were separated by shifting them vertically; unfortunately, the authors of Ref.~\cite{pal2011} were unable to specify the shifts,\footnote{A.N.~Pal, {\it private communication.}} which does not allow accurate comparison of the theoretical and observed noise magnitudes, and in particular, determination of $\delta$). Yet, the similarity of theoretical and experimental shapes of $\varkappa(n),$ evident in Fig.~\ref{fig4}, is quite revealing. Namely, in conjunction with their order-of-magnitude closeness and the possibility to attribute the magnitude difference to $\delta \ne 0,$ it suggests that what is observed in the experiment is actually a manifestation of the quantum indeterminacy, rather than a result of conductance fluctuations. In the absence of consistent theory of conductance $1/f$-noise that would describe both the frequency profile and the noise magnitude, to decide which contribution dominates calls for comprehensive measurements which would allow a sufficiently accurate determination of the frequency exponent to verify the trends predicted by Eq.~(\ref{sf4}).

To summarize, the quantum bound on $1/f$ voltage power spectrum in materials with a conical energy-momentum dispersion of charge carriers is finite despite an apparent zero-mass singularity of the bound in the case of parabolic dispersion \cite{kazakov1}. Rather, the masslessness of charge carriers shows itself as a singularity of the momentum decomposition of the reduced power spectrum at zero particle momentum. This singularity has a directly measurable consequence, namely, it gives rise to a sharp peak in the power spectrum versus charge carrier density at small values thereof. In application to monolayer graphene, this peak becomes M-shaped on account of a continuous transition from the electron to hole conductivity, and this M-shape turns out to be congruent with the experimentally observed.

\pagebreak

\end{document}